\documentstyle[aas2pp4]{article}

\begin{document}

\title{EVOLUTION OF MOLECULAR ABUNDANCE IN PROTOPLANETARY DISKS}
\author{\sc Yuri Aikawa \footnote{present address: National Astronomical
Observatory, Mitaka, Tokyo 181, Japan}}
\affil{Department of Earth and
Planetary Science, University of Tokyo,\\ Bunkyo-ku, Tokyo 113, Japan}
\author{\sc Toyoharu Umebayashi}
\affil{Data Processing Center, Yamagata University, Yamagata 990, Japan}
\author{\sc Takenori Nakano}
\affil{Nobeyama
Radio Observatory, National Astronomical Observatory,\\
Nobeyama, Minamisaku, Nagano 384-13, Japan}
\begin{center}
{\sc and}
\end{center}
\author{\sc Shoken M. Miyama}
\affil{National
Astronomical Observatory, Mitaka, Tokyo 181, Japan}

\begin{abstract}
We investigate the evolution of molecular abundance in quiescent
protoplanetary disks which are presumed to be around weak-line
T Tauri stars.
In the region of surface density less than $10^2$ {\rm g
cm}$^{-2}$ (distance from the star $\gtrsim 10$ AU in
the minimum-mass solar nebula), cosmic rays are barely
attenuated even in the midplane of the disk and
produce chemically active ions such as He$^+$ and H$_{\rm 3}^+$.
Through reactions with these ions CO and N$_2$
are finally transformed into {\rm CO}$_2$, {\rm NH}$_3$, and
{\rm HCN}. In the region where the temperature is low enough for
these products to 
freeze onto grains, considerable amount of carbon and nitrogen is
locked up in the ice mantle and is depleted from the gas phase in a
time scale $\lesssim 3\times 10^6$ yr. Oxidized (CO$_2$) ice and
reduced (NH$_3$ and hydrocarbon) ice naturally coexist in this part of
the disk. The molecular abundance both in the gas phase and in ice
mantle varies significantly with the distance from the central star.

\end{abstract}


\section{INTRODUCTION}

Radio, infrared, and optical observations have recently revealed the
existence of circumstellar disks around young stellar objects (Sargent
\& Beckwith 1994; O'dell \& Wen Zheng 1994). Planet formation must be going
on at least in some of them. Theoretical study on the evolution of
molecular abundance in such `` protoplanetary '' disks is important
from various points of view. First, it will directly show what material
the bodies in the planetary systems are made from.
Secondly, molecular abundance can be
a useful probe in investigating the formation processes of planetary
systems. For example, we may be able to reduce the processes, places,
and epochs of the formation of
primitive bodies, such as comets, by comparing their molecular
composition with the theoretical results on the distribution and
evolution of molecular abundance in the disk. Thirdly, study of the
structure and evolution of the disk by observations of molecular lines
should ask for a help of theoretical study on molecular evolution. 
For example, the amount of the gaseous
component in the disk is estimated by observations of molecules other
than the main component H$_2$ because H$_2$ can hardly emit photons
except at very inner region of the disk. Knowledge
on the abundance of the molecules relative to hydrogen is
indispensable for such study.

Some theoretical works have been done on the molecular abundance in
protoplanetary disks.
Prinn and his colleagues (Prinn 1993, and references therein)
investigated the molecular evolution in
fully turbulent accretion disks and proposed the kinetic
inhibition (KI) model.
Their basic idea is that
the molecular abundance in the matter flowing outwards
is `` quenched '' when its temperature has decreased to a value below
which the time scale for chemical
reaction is larger than the dynamical time scale.
Adopting the lifetime of the disk $\sim 10^{13}$ s, they concluded that
the molecular abundance was quenched at
$850-1500$K, and that CO and N$_2$ were the dominant component
in the solar nebula. They considered neutral-neutral
reactions which can be efficient in the region of $T
\gtrsim$ several $\times 10^2$ K.

In this letter we investigate the evolution of molecular abundance in a
relatively quiescent disk in the post-accretion phase, the so-called
weak-line T Tauri (WTTS)
phase and thereafter, in which transport and mixing
of matter is not so efficient as in the earlier active phases mentioned above. 
We consider the disk regions where the column density is lower than
the attenuation length of cosmic rays, 96 g cm$^{-2}$ (Umebayashi \&
Nakano 1981). The
minimum-mass solar nebula (Hayashi 1981)  satisfies this condition in
the region more than about 7AU away from the central star with
the temperature $T\lesssim 10^2$ K. We take into account the ionization by
cosmic rays and the subsequent
ion-molecule reactions, which were not considered in the previous works.

\section{DISK MODEL AND REACTION NETWORK}

 As the model of the protoplanetary disk,
we adopt the so-called minimum-mass solar nebula (Hayashi 1981),
which has the the distribution of the surface density
$\Sigma (R) = 54~ \left(R/10 {\rm AU}\right)^{-3/2} {\rm g~ 
cm^{-2}}$ and temperature $T(R)=89 \left(R/10{\rm
AU}\right)^{-1/2}{\rm K}$, where $R$ is the distance from the central star.
This model is consistent with the
distribution of the temperature and the surface density
estimated from the observations of dust continuum (Beckwith et al. 1990).
Hydrostatic equilibrium determines the density distribution
perpendicular to the midplane of the disk (e.g., Aikawa et al. 1996).

We consider the gas-phase reactions, adsorption onto grains, and
thermal desorption from grains.
The reaction network we use is essentially the
same as the one described in Aikawa et al. (1996). We adopt the
UMIST94 data base (Millar et al. 1991; Farquhar \& Millar 1993)
for the gas-phase reactions, and also take into account some
three-body reactions
referring to Brasseur \& Solomon (1986).
Table 1 shows all the atoms and molecules included in our network
(except H, H$_2$, and He) and their adsorption energies.
Because the attenuation of cosmic rays is inefficient, we adopt the
ionization rate in the dark cloud, $\zeta \approx 10^{-17}$ s$^{-1}$. We
neglect the ionization and dissociation by the interstellar and stellar
ultraviolet radiation since it can be attenuated by grains in a thin
surface layer of the disk (see $\S$ 4.3). We take the
sticking probability $S=0.3$ for the collision
of a gas particle with a grain (Williams 1993).
For simplicity, we assume that all grains are spherical with
radius $a = 10^{-5}$ cm
and that the dust-gas ratio is the same as that in the interstellar
clouds. Although the dust-gas ratio near the midplane
might be enhanced by
sedimentation of dust, it does not affect much the results;
we have obtained almost the same results with the dust-gas
ratio higher by orders of magnitude.

Since our reaction network consists of one- and two-body
reactions and contains only a small number of three-body reactions,
we consider the region of the density by number of hydrogen nuclei,
$n_{\rm H}\lesssim
10^{12}$cm$^{-3}$, which corresponds to $R\gtrsim 10$ AU for the
minimum-mass solar nebula.
With a typical rate coefficient $k \sim 10^{-30}$cm$^6$s$^{-1}$ for
the three-body reactions we have in this density range
$kn$(H$_2$)$\lesssim 10^{-18}$cm$^3$s$^{-1}$ even for the most
abundant species H$_2$ as a third body.
This is smaller than the characteristic rate coefficient
$\sim 10^{-17}$cm$^3$s$^{-1}$ for radiative association, which is the
slowest among the two-body reactions. Thus
the three-body reactions are not important
in this density range.

For simplicity and clarity, we assume that the density
and temperature do not change with time. Since we are mainly
interested in the quiescent (WTTS) phase, we neglect the radial transport
of matter and perform calculation up to $\sim 10^7$ yr.
The initial molecular abundance is the same
as in Aikawa et al. (1996), which assumed mostly the observed
abundance in the dark cloud TMC-1. However, the molecular abundance
may change during the active phase of the disk.
To find out the influence of the initial condition, we also perform
calculation with another set of the initial abundance; all carbon is in the
form of CO, the remaining oxygen is locked up in
water ice mantle, and nitrogen is in N$_2$.

\section{NUMERICAL RESULTS}

Since the result is found to be much more sensitive to the
temperature than to the density, we show as the representative
cases the results at $T=30$
and 90K corresponding to the regions of
$R=87$ and 9.7AU with the density $n_{\rm H}$ at the midplane $2.9\times 10^9$
and $1.2\times 10^{12}$ cm$^{-3}$, respectively.

Figures $1a$ and $1b$ show the results for the first set of the
initial condition.
The solid lines show the time variation of the abundance
of carbon-bearing molecules relative to hydrogen.
In the region of $T=30$K (Figure 1$a$) the abundance of CO
remains nearly constant at its initial value $n({\rm CO})/n_{\rm H}
\approx 7\times 10^{-5}$ up to $10^6$yr. Thereafter
CO decreases steeply, and CO$_2$ ice and HCN ice become dominant.
The main process of forming CO$_2$ and HCN is as follows.
Through the reaction with H$_3^+$, CO is transformed
into HCO$^+$, some of which become HCO mainly through
the grain surface recombination.
Reactions of HCO with C and N finally form CO$_2$ and
HCN in the gas phase\footnote{By recombination on grains HCO$^+$ is
transformed into HCO radical or CO + H with
unknown branching ratio. We assumed that the former branch is
dominant. The time scale of the formation of
CO$_2$ and HCN becomes longer if we assume otherwise.}.
Once CO$_2$ and HCN are formed, they are
adsorbed by grains in a time as short as $\sim
10$ yr (Aikawa et al. 1996).
Since these molecules are hardly desorbed at $T=30$K, they are locked up
in the ice mantle, and as a result carbon is depleted from
the gas phase.
At somewhat higher temperatures (e.g., $T=90$K as in Figure 1$b$),
the thermal desorption is
efficient enough to compensate the adsorption; although CO$_2$
becomes abundant, CO is also abundant on a
time scale of $10^7$ yr.

The dotted lines in Figures 1$a$ and 1$b$ show the time variation
of the abundance of oxygen-bearing molecules.
Oxygen is initially assumed to be mostly in the form of O atom and CO
molecule.
The abundance of O$_2$ in the gas phase and H$_2$O
ice in the grain mantle increases extensively with time.
The H$_2$O molecules form mainly in the following process.
The reaction between O and H$_3^+$ produces
OH$^+$, most of
which are transformed into H$_3$O$^+$ through H atom-abstractions from
H$_2$. Then
H$_3$O$^+$ recombines to produce H$_2$O.
In the later stages where O$_2$ becomes dominant, the main reaction
of forming OH$^+$ is the H atom-abstraction by O$^+$,
which is produced via destruction of O$_2$ by He$^+$.
Since H$_2$O freezes onto grains at these temperatures
while O$_2$ does not, most oxygen freezes out in the
form of H$_2$O ice and the abundance of O$_2$ decreases in a
time scale of several $\times 10^6$ yr.

The dashed lines in Figure 1 show the time variation of the abundance
of nitrogen-bearing molecules.
Initially most nitrogen is assumed to be in the form of N
atoms. As time goes on, N$_2$, HCN ice, and NH$_3$ ice become dominant.
The main formation process of NH$_3$ is as follows.
The reaction of N$_2$ with He$^+$ forms N$^+$, which is transformed
into NH$_4^+$ by repeating H atom-abstraction. Finally NH$_4^+$ is
turned into NH$_3$ through dissociative recombination and grain-surface
recombination.
At temperatures
below about 70K, NH$_3$ freezes onto grains
and most nitrogen is depleted from the gas phase.
At higher temperatures as in Figure $1b$, NH$_3$ cannot freeze
and N$_2$ is dominant even at later stages.

\section{DISCUSSION}

\subsection{\em Implications to Planetary Science}

We review the main points of our results and discuss
the implications to planetary science. First, we have found that
CO and N$_2$ are transformed into CO$_2$
and NH$_3$. The
main cause for this is in cosmic rays, which have not been taken into
account in the previous works. Cosmic rays
produce chemically active ions such as H$_3^+$ and He$^+$ which can
destroy CO and N$_2$. This results in the
depletion of CO in the gas phase and hence the faintness of
its emission lines in the cold regions ($T\lesssim 70$K) of
the disks at the age $\gtrsim$ several $\times 10^6$ yr.
This also means that CO$_2$ ice
and NH$_3$ ice will be detected in these regions.
Secondly, our results show that oxidized (CO$_2$) ice and reduced
(NH$_3$ and hydrocarbon) ice coexist in the protoplanetary disks.
This coexistence has also been suggested by the KI model
(Prinn 1993).
This result is to be noted
because their coexistence is one of the most important
characteristics of comets (Yamamoto 1991). CO and N$_2$ would also
coexist in ice mantle if they are physically trapped in water ice
(Bar-Nun et al. 1985),
although this effect is not included in our calculation.
Finally, we have found that the molecular abundance in the gas
and in ice mantle varies considerably with the distance from the
central star. Comparison of this result with the molecular abundance
of comets will enable us to deduce where comets formed,
although quantitative discussion is out of the scope of this paper.

\subsection{\em Dependence on the Initial Condition}

The initial condition we have adopted so far is that considerable amount of
carbon, oxygen, and nitrogen are atomic (see $\S$ 2 in Aikawa et al. 1996).
In order to check the dependence on the initial condition,
we have made calculation for another set of the
initial abundance mentioned in $\S$ 2.
Figure 2 shows the result for this case in the region of $T=30$K.
We can see that CO$_2$ ice
becomes dominant in this case as well. Although there are no
oxygen atoms initially, they are extracted from CO by the
reaction with He$^+$.
The remaining carbon is transformed into hydrocarbon. At higher
temperatures ($T\gtrsim 70$K), CO$_2$ becomes abundant at later
stages, although its abundance
$n_{\rm CO2}/n_{\rm H}=10^{-5}-10^{-6}$ is lower than that for the
first case on the initial condition.
Similarly, N$_2$ is destroyed by He$^+$ and is transformed into HCN ice
and NH$_3$ ice.

\subsection{\em Discussion on the Assumptions}

Although we have neglected the radial transport of matter, our
results are qualitatively applicable to the disks with radial transport
such as the accretion disk and weakly-turbulent disk,
because the chemical paths described in $\S$ 3 are always
active and dominant in the reaction network as long as the matter is in
the region of low temperature and low density ($R\gtrsim 10$AU in the
minimum-mass solar nebula). The molecular evolution in the accretion
disks will be described in the forthcoming paper in more detail.
On the other hand, our results are not applicable to the
disks with efficient mixing where the strong turbulence or
convection mixes matter between the inner hot
($T\gtrsim$ several$\times 100$K)
region and the outer cold region ($R\gtrsim 10$AU) in a time scale $\lesssim
10^6$yr, as investigated by Prinn (1993).

We have neglected the UV photolysis. Smaller grains sink more slowly
toward the midplane of the disk (Nakagawa, Nakazawa, \& Hayashi 1981).
In addition, even a weak turbulence
which cannot efficiently transport matter radially would be sufficient to
prevent the sedimentation of small grains. Therefore, it would be possible
that small grains shield the main part of the disk from the UV
radiation. Mid- to far-infrared excesses observed in the spectra of
WTTS strongly suggest the existence of small
grains at $R \gtrsim 10$AU even in the relatively quiescent disks
(Strom et al. 1989). 

If almost all grains sank to the midplane in less than about $10^6$yr despite
the above discussion and the mid- to far-infrared excesses in WTTS, our
results were not applicable.
The molecular abundance must then be affected by the
photolysis because the line shielding is not efficient enough for
some kinds of species.

\acknowledgements
We would like to thank Drs. T. Yamamoto, Y. Abe, and J. Watanabe
for helpful discussion. Anonymous referee's comments were useful in
improving the manuscript. The numerical calculations were performed partly
at the Astronomical Data Analysis Center of National Astronomical
Observatory, Japan. This work is supported by Grant-in-Aid for
Scientific Research (08640319) of the Ministry of Education, Science, Sports,
and Culture, Japan. Y. A. would like to acknowledge support by JSPS
Research Fellowships for Young Scientists.

\newpage

\begin{deluxetable}{l r l r}
\small
\tablecaption{THE ADSORPTION ENERGIES OF ATOMS AND MOLECULES IN
KELVIN \tablenotemark{a}}
\tablewidth{0pt}
\tablehead{
\colhead{Species} & \colhead{ Adsorption} & \colhead{Species}
& \colhead{Adsorption}\nl
 & \colhead{Energy (K)} &
& \colhead{Energy (K)}}
\startdata
C & 800 ~ ~ ~ ~ &~ C$_2$H & 1460 ~ ~ ~ ~ \nl
N & 800 ~ ~ ~ ~ &~ C$_3$ & 2010 ~ ~ ~ ~ \nl
O & 800 ~ ~ ~ ~ &~ CCN & 2010 ~ ~ ~ ~ \nl
Na& 11800 ~ ~ ~ ~ &~ CO$_2$ & 2690\tablenotemark{b} ~ ~ ~ ~ \nl
Mg & 5300 ~ ~ ~ ~ &~ NH$_2$ & 860 ~ ~ ~ ~ \nl
Si & 2700 ~ ~ ~ ~ &~ NO$_2$ & 2520 ~ ~ ~ ~ \nl
S & 1100 ~ ~ ~ ~ &~ OCN & 2010 ~ ~ ~ ~ \nl
Fe & 4200 ~ ~ ~ ~ &~ OCS & 3000 ~ ~ ~ ~ \nl
HS & 1500 ~ ~ ~ ~ &~ O$_2$H & 1510 ~ ~ ~ ~ \nl
CH & 650 ~ ~ ~ ~ &~ O$_3$ & 2520 ~ ~ ~ ~ \nl
C$_2$ & 1210 ~ ~ ~ ~ &~ SO$_2$ & 3460\tablenotemark{b} ~ ~ ~ ~ \nl
CN & 1510 ~ ~ ~ ~ &~ H$_2$CO & 1760 ~ ~ ~ ~ \nl
CO & 960\tablenotemark{b} ~ ~ ~ ~ &~ H$_2$CS & 2250 ~ ~ ~ ~ \nl
CS & 2000 ~ ~ ~ ~ &~ CH$_3$ & 1160 ~ ~ ~ ~ \nl
NH & 600 ~ ~ ~ ~ &~ C$_2$H$_2$ &2400\tablenotemark{c} ~ ~ ~ ~ \nl
N$_2$ & 710\tablenotemark{c} ~ ~ ~ ~ &~ C$_3$H & 2270 ~ ~ ~ ~ \nl
NO & 1210 ~ ~ ~ ~ &~ C$_3$N & 2720 ~ ~ ~ ~ \nl
NS & 2000 ~ ~ ~ ~ &~ NH$_3$ & 3080\tablenotemark{b} ~ ~ ~ ~ \nl
OH & 1260 ~ ~ ~ ~ &~ H$_2$C$_3$ & 2110 ~ ~ ~ ~ \nl
O$_2$ & 1210 ~ ~ ~ ~ &~ HC$_3$N &2970 ~ ~ ~ ~ \nl
SiH & 2940 ~ ~ ~ ~ &~ CH$_4$ & 1080\tablenotemark{c} ~ ~ ~ ~ \nl
SiC & 3500 ~ ~ ~ ~ &~ CH$_3$O & 1710  ~ ~ ~ ~ \nl
SiO & 3500 ~ ~ ~ ~ &~ C$_3$H$_2$ & 2110  ~ ~ ~ ~ \nl
SiS & 3800 ~ ~ ~ ~  &~ CH$_3$CN &2270 ~ ~ ~ ~ \nl
SO & 2000 ~ ~ ~ ~ &~ CH$_3$OH &4240\tablenotemark{b} ~ ~ ~ ~ \nl
H$_2$O & 4820\tablenotemark{b} ~ ~ ~ ~ &~ CH$_3$O$_2$ & 2370  ~ ~ ~ ~ \nl
H$_2$S & 1800 ~ ~ ~ ~ &~ C$_3$H$_3$ & 2220 ~ ~ ~ ~ \nl
HCN & 4170\tablenotemark{c} ~ ~ ~ ~ &~ H$_3$C$_3$N & 3270  ~ ~ ~ ~ \nl
HCO & 1510 ~ ~ ~ ~ &~ C$_3$H$_4$ & 2470 ~ ~ ~ ~ \nl
CH$_2$ & 960 ~ ~ ~ ~ &~ CH$_3$OOH & 2620 ~ ~ ~ ~ \nl
\enddata
\tablenotetext{a}{Allen \& Robinson 1977 and Hasegawa \& Herbst 1993
except for $^b$ and $^c$}
\tablenotetext{b}{Sandford \& Allamandola 1993}
\tablenotetext{c}{Yamamoto, Nakagawa, \& Fukui 1983}
\end{deluxetable}

\clearpage

\figcaption[]{Evolution of the molecular abundance in two
representative regions of the disk, (a) $R=87$AU
($n_{\rm H}=2.9\times 10^9$ cm$^{-3}, T=30$K) and (b) $R=9.7$AU
($n_{\rm H}=1.2\times 10^{12}$ cm$^{-3}, T=90$K). The initial
molecular abundance was determined referring to the abundance in
dark clouds. The solid, dotted, and dashed lines represent
the abundance of carbon-, oxygen-, and nitrogen-bearing
molecules, respectively, relative to hydrogen.}

\figcaption[]{Evolution of the molecular abundance at $R=87$AU
($n_{\rm H}=2.9\times 10^9$ cm$^{-3}, T=30$K) with the initial
condition that carbon is in the form of CO, nitrogen is in the form of
N$_2$, and the remaining oxygen is locked up in water ice mantle.
The other details are the same as in Fig. 1}

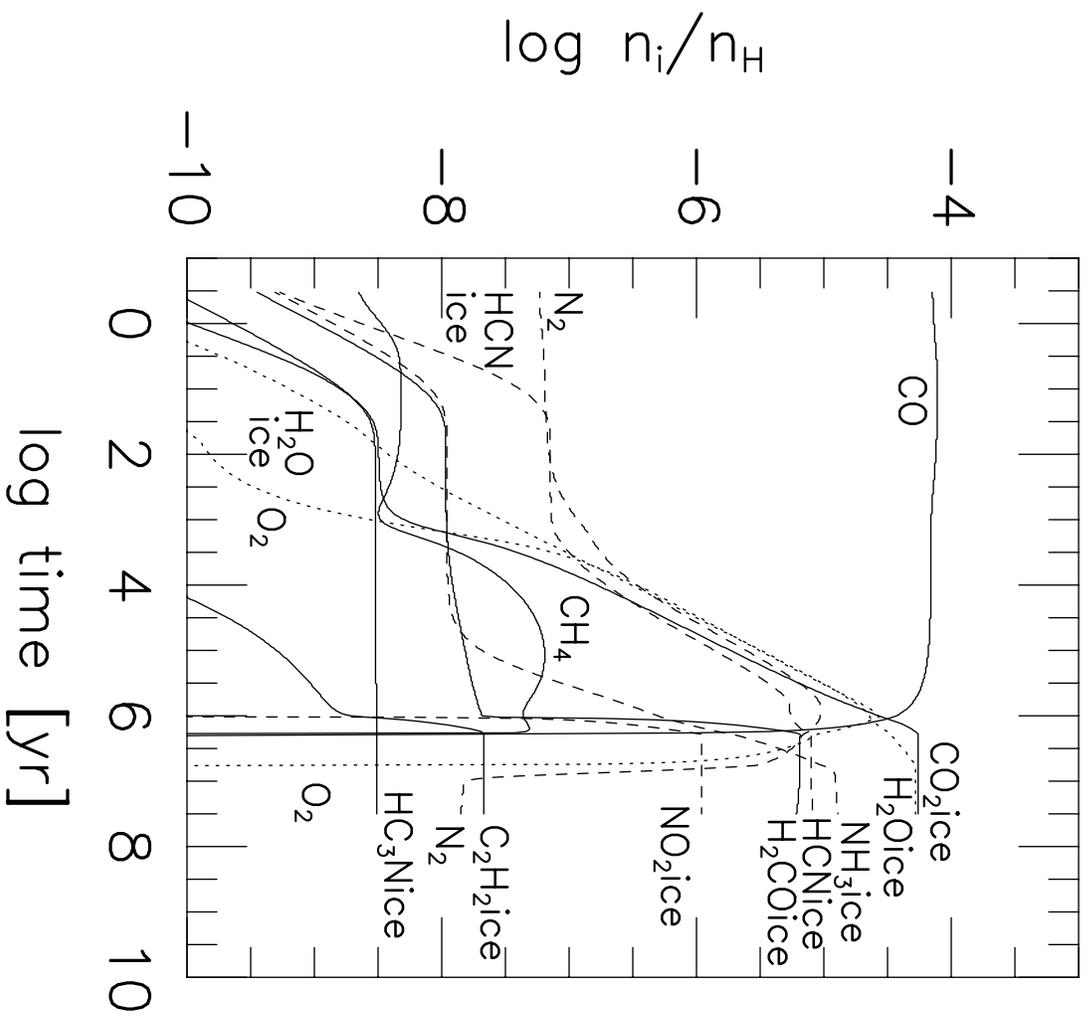
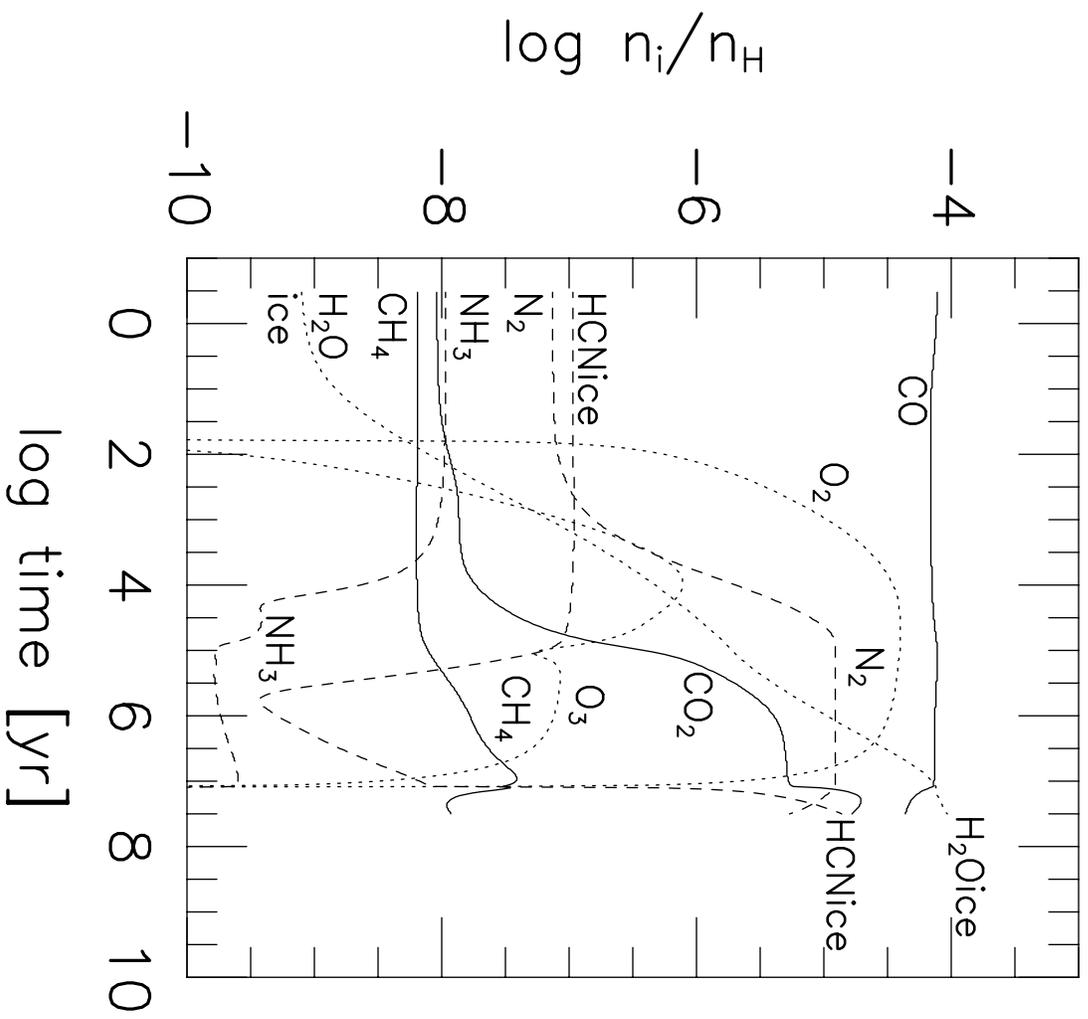
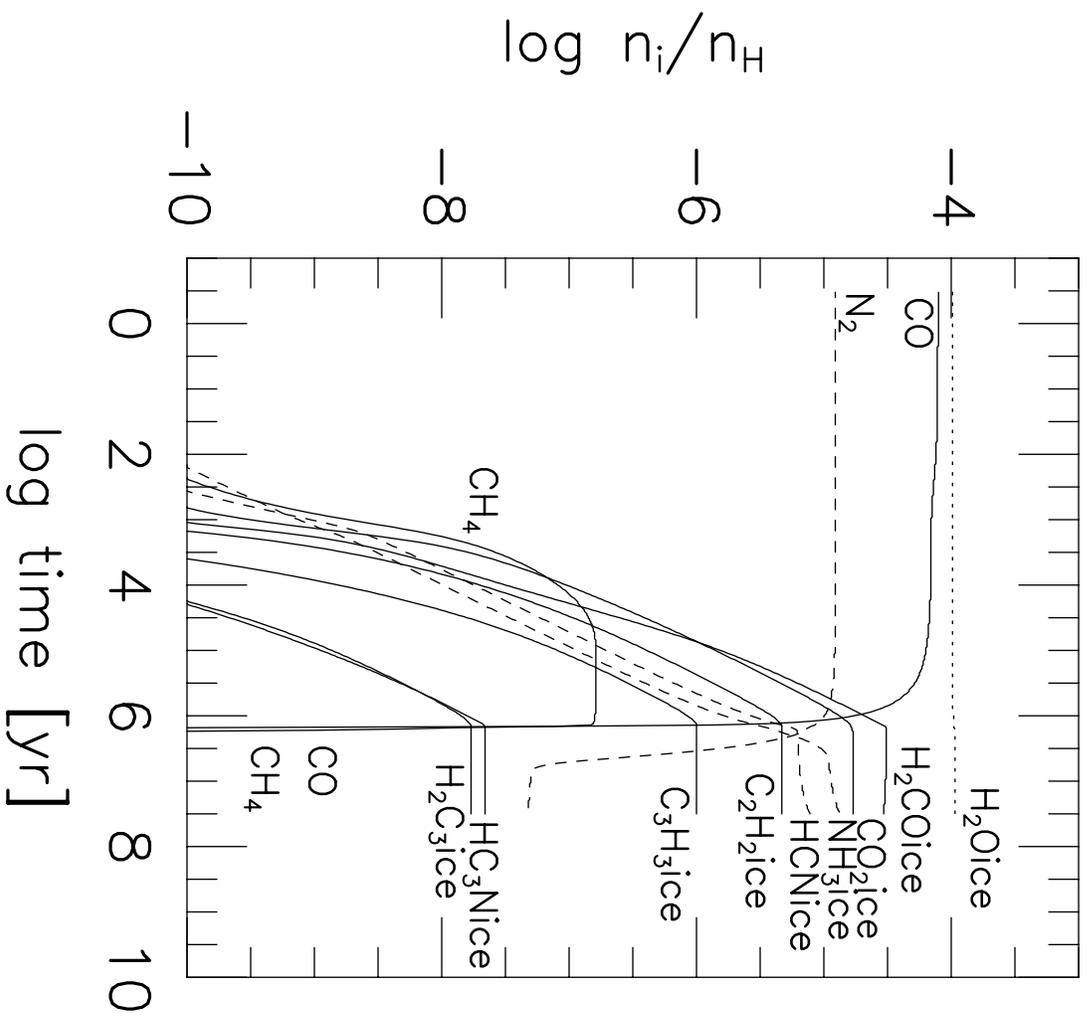
\end{document}